\documentclass[9pt,twocolumn,twoside,dvipsnames]{osajnl}

\journal{josab}

\setboolean{shortarticle}{false}

\usepackage{amsmath}

\usepackage{physics}

\usepackage{xfrac}

\usepackage[mathscr]{euscript}

\usepackage{textgreek}

\usepackage{xcolor}
\usepackage{tikz,pgfplots}
\usetikzlibrary{3d,pgfplots.groupplots} 
\usepgfplotslibrary{units} 
\usepgfplotslibrary{external} 
\pgfplotsset{compat=newest, legend image code/.code={\draw[mark repeat=2,mark phase=2] plot coordinates {(0cm,0cm) (0.15cm,0cm) (0.3cm,0cm)};}}

\definecolor{col_red}{RGB}{255,0,0}
\definecolor{col_orange}{RGB}{255,102,0}
\definecolor{col_yellow}{RGB}{255,204,0}
\definecolor{col_green}{RGB}{0,128,0}
\definecolor{col_blue}{RGB}{0,0,255}
\definecolor{col_purple}{RGB}{128,0,128}
\definecolor{col_grey}{RGB}{150,150,150}
\definecolor{col_grey_2}{RGB}{128,128,128}
\definecolor{col_light_green}{RGB}{0,204,0}

\pgfdeclareverticalshading{rainbow_1}{100bp}
{color(0bp)=(red); color(40bp)=(red); color(44bp)=(yellow); color(48bp)=(green); color(52bp)=(cyan); color(56bp)=(blue); color(60bp)=(violet); color(100bp)=(violet)}
\pgfdeclareverticalshading{rainbow_2}{100bp}
{color(0bp)=(red); color(45bp)=(red); color(46bp)=(yellow); color(47bp)=(green); color(48bp)=(cyan); color(49bp)=(blue); color(50bp)=(violet); color(100bp)=(violet)}
\pgfdeclareverticalshading{rainbow_3}{100bp}
{color(0bp)=(red); color(25bp)=(red); color(35bp)=(yellow); color(45bp)=(green); color(55bp)=(cyan); color(65bp)=(blue); color(75bp)=(violet); color(100bp)=(violet)}

\title{Optimising hybrid rotational femtosecond/picosecond coherent anti-Stokes Raman spectroscopy (HR-CARS) in nitrogen at high pressures and temperatures}

\author[1,*]{Nils Torge Mecker}
\author[2]{Trevor L. Courtney}
\author[2]{Brian D. Patterson}
\author[1]{David Escofet-Martin}
\author[1]{Brian Peterson}
\author[2]{Christopher J. Kliewer}
\author[1]{Mark Linne}

\affil[1]{School of Engineering, The University of Edinburgh, Edinburgh EH8 3JL, Scotland}
\affil[2]{Combustion Research Facility, Sandia National Laboratories, Livermore, CA 94551, USA}
\affil[*]{Corresponding author: n.mecker@ed.ac.uk}

\begin{abstract}
We demonstrate the use of hybrid rotational femtosecond/picosecond (fs/ps) coherent anti-Stokes Raman spectroscopy (HR-CARS) as a technique for temperature measurements in nitrogen gas at high pressures and temperatures. A broadband pulse shaper-adjusted 42 fs pulse interacts with a narrow-bandwidth, frequency-upconverted 5.5 ps pulse in a cell containing N\textsubscript{2} at pressures of 1-70 atm and temperatures of 300-1000 K. A computational code is used to model spectra and fit experimental results to obtain best-fit temperatures. We demonstrate good qualitative fits as well as good accuracy and precision between thermocouple measured and best-fit temperatures over the explored pressure and temperature regimes. The overall average percentage temperature difference between thermocouple measurements and best-fit temperatures is -0.3\% with a standard deviation of 7.1\%, showing the suitability of HR-CARS for characterising high pressure and temperature environments.
\end{abstract}

\setboolean{displaycopyright}{true}

\begin{document}

\maketitle

\section{Introduction}
Coherent anti-Stokes Raman spectroscopy (CARS) was first used for gas-phase measurements in the 1970s \cite{regnier_1973}. Since then, it has become a widely applied technique for measurements of temperature and species concentration in flames and other gas-phase applications \cite{eckbreth_book_1996}. CARS is a four-wave mixing technique that combines three laser beams to create a fourth output beam which carries the spectral information. It has several advantages over other methods such as laser-induced fluorescence (LIF). These advantages include very good spectral resolution and a laser-like directional signal \cite{roy_2010}. CARS has been applied to a variety of combustion processes at elevated temperatures and pressures, both as vibrational CARS (VCARS) and also as pure rotational CARS (RCARS). VCARS experiments in nitrogen (N\textsubscript{2}) have been reported for pressures $\leqslant$2500 atm, including temperature models and observations on pressure-induced collisional narrowing \cite{bouche_1990, dreier_1994}. VCARS, however, suffers from a collapse of the rotational structure at higher pressures, as the closely spaced ro-vibrational lines start to mix \cite{hall_1980}. In RCARS the line spacing is much bigger, resulting in more highly resolved spectra, even at higher pressures. Studies of RCARS on both N\textsubscript{2} and O\textsubscript{2} at various temperatures and pressures up to 150 atm have been reported \cite{bood_2000, knopp_2002, seeger_2003, vestin_2007}.

The traditionally used nanosecond CARS has some important limitations, including interference from the nonresonant (NR) background signal, low acquisition rates and dependence on collisional dephasing. Picosecond CARS overcomes interference from the nonresonant background as the probe beam can be delayed in time from the pump and Stokes beams to acquire signal at a time when the NR contribution has decayed significantly, while the CARS signal remains strong. Unfortunately, collisional dephasing and low acquisition rates remain limitations of picosecond CARS, similar to nanosecond CARS. These problems were overcome by using femtosecond CARS, a technique that enables collision-free measurements at high repetition rates. However, while ns and ps CARS achieve high spectral resolution, fs CARS measures spectra in the time domain, as the pulses are very short (10-100 fs), but accordingly broad in the spectral domain. In 2006, hybrid femtosecond/picosecond (fs/ps) CARS was developed by Prince \textit{et al.} \cite{Prince_2006}. By using broadband femtosecond laser pulses to create the Raman coherences and by probing them with a narrow-band picosecond pulse, it combines the advantages of ps and fs CARS. The NR background is avoided, while high repetition rates and high peak power can be achieved, making single shot frequency domain measurements possible. In addition, collisional dephasing can be avoided, at least at lower pressures. For example, at 300 K and 60 atm, the collisional dephasing time is about 15 ps. In the two-beam hybrid fs/ps RCARS approach, a single femtosecond beam is used as the pump and Stokes beams to excite the rotational Raman coherences. This provides even further advantages, as it simplifies the experimental setup and reduces the effect of beam steering, which is often encountered in turbulent conditions \cite{bohlin_2013}.

Short-pulse CARS techniques have only been tested in a few experiments at higher pressures. Papers have been published for moderate pressures at room temperature in the time domain and up to 900 K in the frequency domain, using fs-CARS and hybrid fs/ps CARS \cite{miller_2011_a, wrzesinski_2013, matthaus_2016, kerstan_2017, kearney_2015}. In this paper we present an experimental setup for two-beam hybrid rotational fs/ps CARS (HR-CARS) of N\textsubscript{2} at pressures up to 70 atm and temperatures up to 1000 K. The research presented here is a significant extension of a previous paper \cite{courtney_2019_a}, including a detailed presentation over the entire range of pressures and temperatures. Our setup includes second harmonic bandwidth compression (SHBC) to create the picosecond probe beam \cite{courtney_2019_b} and a pulse shaper to create a transform-limited (TL) femtosecond pump/Stokes beam at the interaction volume. Further, we present a computational model to generate theoretical HR-CARS spectra and we validate it by fitting the experimental spectra to obtain best-fit temperatures, thereby showing the usefulness of this technique and potential use for thermometry in high pressure combustion environments in the future.

\section{Theory and Model}
Quantum mechanics forms the theoretical basis of CARS spectroscopy. In a semi-classical treatment, using the density matrix and the Schr\"{o}dinger equation together with a classical electromagnetic optical wave, the 3\textsuperscript{rd} order time-domain polarisation, upon which the CARS signal depends, can be derived \cite{mukamel_book_1995, hamm_book_2005}:
\begin{align}
P^{(3)}_{CARS}(t) = &\left(\frac{i}{\hbar}\right)^3 \int_{0}^{\infty} dt_3 \int_{0}^{\infty} dt_2 \int_{0}^{\infty} dt_1 R_4(t_3,t_2,t_1)\nonumber\\
&E_3(t-t_3)E_2(t-t_3-t_2)E_1(t-t_3-t_2-t_1)
\label{eq:3rd_order_pol_general}
\end{align}

\begin{figure}[htbp]
\centering
\resizebox{0.75\columnwidth}{!}{%
\includegraphics{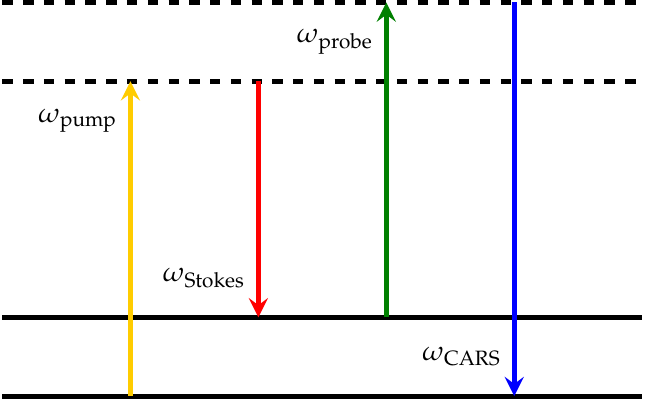}}
\caption{Schematic diagram of the four-wave mixing CARS process, showing the interaction between the pump, Stokes and probe photons to produce a CARS signal photon. Solid lines represent real molecular states (rotational states in the case of RCARS) and dashed lines represent virtual states.}
\label{fig:CARS_process}
\end{figure}

$E_1$, $E_2$ and $E_3$ are the time domain electric fields of the pump, Stokes and probe beams respectively. $R_4$ is the 3\textsuperscript{rd} order molecular response function and $t_1$, $t_2$ and $t_3$ represent coherence timescales. This nonlinear four-wave mixing process is depicted in Figure \ref{fig:CARS_process}.

The 3\textsuperscript{rd} order polarisation can be simplified, by assuming that the molecular response is fast compared to the pump and probe pulse coherence time scales, $t_1$ and $t_3$. This assumption is valid as the pump and probe pulses are detuned from any direct one-photon electronic transition, so that molecular dephasing over the $t_1$ and $t_3$ timescales is almost instantaneous. The 3\textsuperscript{rd} order response function can then be replaced by $R_4(t_3,t_2,t_1) \rightarrow \delta(t_3)R_4(0,t_2,0)\delta(t_1)$ \cite{miller_phd_2012, stauffer_2014}. A further simplification is the assumption that the pump and Stokes pulses are short and impulsive compared to the probe pulse, so that the pump and Stokes electric fields can be replaced by delta functions. This is a valid assumption for HR-CARS, as the picosecond probe is about two orders of magnitude longer than the femtosecond pump/Stokes beam. Under these assumptions, the 3\textsuperscript{rd} order polarisation is given by:
\begin{align}
P^{(3)}_{CARS}(t,\tau) = \left(\frac{i}{\hbar}\right)^3 E_{pr}(t-\tau) R_{CARS}(t) e^{i\omega_{pr}(t-\tau)}
\label{eq:3rd_order_pol_specific}
\end{align}

\noindent where $E_{pr}$ is the electric field envelope of the probe beam with centre frequency $\omega_{pr}$ and time delay $\tau$ relative to the pump/Stokes beams. The molecular response function is denoted as $R_{CARS}(t)$ now. The 3\textsuperscript{rd} order polarisation in the frequency-domain is given by a Fourier transform of Equation \ref{eq:3rd_order_pol_specific}. The frequency-domain CARS signal simply is the square of the absolute value of the frequency domain 3\textsuperscript{rd} order polarisation:
\begin{align}
P^{(3)}_{CARS}(\omega,\tau) = \mathscr{F} \left[P^{(3)}_{CARS}(t,\tau)\right]
\label{eq:fourier_transform}
\end{align}
\begin{align}
I_{CARS}(\omega,\tau) \propto \abs{P^{(3)}_{CARS}(\omega,\tau)}^2
\label{eq:cars_signal}
\end{align}

\begin{figure}[htbp]
\centering
\includegraphics{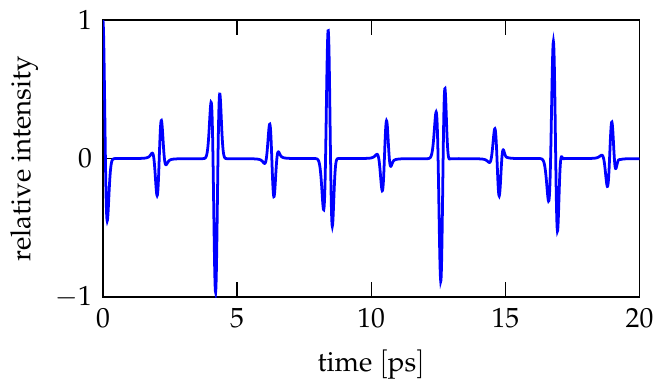}
\caption{Real part of the molecular response function of N\textsubscript{2} in the time domain for pure rotational S-branch transitions at 293 K and 1 atm. The first full revival is at 8.4 ps (and the second at 16.7 ps). Partial rephasing occurs at $\sfrac{1}{4}\ \tau_{full}$, $\sfrac{1}{2}\ \tau_{full}$ and $\sfrac{3}{4}\ \tau_{full}$.}
\label{fig:N2_molecular_response}
\end{figure}

Equations \ref{eq:3rd_order_pol_specific}, \ref{eq:fourier_transform} and \ref{eq:cars_signal} are the underlying mathematics used in our computational code. The 3\textsuperscript{rd} order molecular response function, $R_{CARS}(t)$, is given by the following phenomenological expression for S-branch rotational transitions \cite{miller_phd_2012}:
\begin{align}
R_{CARS}(t) = &\sum_{v}\sum_{J} I_{v,J;v,J+2}\nonumber\\
&\times exp \left[\frac{t}{\hbar} \left(i\Delta E_{v,J;v,J+2} - \frac{1}{2} \Gamma^S_{v,J;v,J+2}\right)\right]
\label{eq:molecular_response}
\end{align}

\noindent where the sum is performed over rotational states $J$ at each vibrational level $v$. $\Delta E_{v,J;v,J+2}$ and $\Gamma^S_{v,J;v,J+2}$ are the frequency and linewidth of a transition between rotational states $J$ and $J+2$ at vibrational level $v$ respectively. $I_{v,J;v,J+2}$ is the corresponding Raman transition strength. For each individual transition, Equation \ref{eq:molecular_response} is a sinusoidal curve with a frequency based on the transition frequency ($\Delta E_{v,J;v,J+2}$) which decays exponentially with linewidth ($\Gamma^S_{v,J;v,J+2}$) and has an amplitude given by the transition strength ($I_{v,J;v,J+2}$). The sum over all rotational states at each vibrational level results in a time-domain Raman signal. The individual exponentially decaying sinusoidal waves are in phase at certain times and out of phase at others, resulting in a periodic structure as shown in Figure \ref{fig:N2_molecular_response}. This is because as the rotational coherences set up by the pump/Stokes beams evolve in time they become dephased and then rephased again in certain intervals. Whenever they are rephased there is a peak in the Raman signal. For rotations, rephasing occurs in intervals of $\tau_{full} = \sfrac{1}{(2B_e c)}$, where $c$ is the speed of light and $B_e$ the equilibrium rotational constant. The first full revival of N\textsubscript{2} ($B_e$=1.99826 cm\textsuperscript{-1} \cite{palmer_carsft_1989}) is at 8.4 ps. Partial rephasing occurs at $\sfrac{1}{4}\ \tau_{full}$, $\sfrac{1}{2}\ \tau_{full}$ and $\sfrac{3}{4}\ \tau_{full}$.

The Raman transition strength is given by the following \cite{miller_phd_2012}:
\begin{align}
I_{v,J;v,J+2} \propto \frac{4}{45} b_{J,J+2} \times (\gamma'_v)^2 \times F_{rot}(J) \times \Delta\rho_{v,J;v,J+2}
\label{eq:transition_strength}
\end{align}

\noindent where $b_{J,J+2}$ is the Placzek-Teller Coefficient, $\gamma'_v$ is the polarisation anisotropy, $F_{rot}(J)$ is the Herman-Wallis Factor and $\Delta\rho_{v,J;v,J+2}$ is a population difference factor, based on the Boltzmann distribution of the rotational states.

The Placzek-Teller Coefficient, first described by Placzek and Teller in 1933 \cite{placzek_1933}, determines the dependence of the Raman transition strength on the rotational level. As a molecule rotates faster (i.e. as its rotational quantum number increases), its internuclear separation gets larger, changing its moment of inertia. This centrifugal distortion changes the polarisability of the molecule, which in turn changes the Raman transition strength. For S-branch transitions of linear diatomic molecules, this factor is:
\begin{align}
b_{J,J+2} = \frac{3(J+1)(J+2)}{2(2J+1)(2J+3)}
\label{eq:PT_coefficient}
\end{align}

Similar to the change in polarisability (and hence the Raman transition strength) due to the rotation of the molecule, there is also a change in polarisability due to the interaction between rotations and vibrations. This interaction arises from there being many vibrations during a single rotation. The change in Raman transition strength due to this ro-vibrational interaction is described by the Herman-Wallis Factor. For S-branch rotational transitions of nitrogen, there are different versions of the Herman-Wallis Factor in the literature \cite{bohlin_2012}. Our model uses the S-branch Herman-Wallis Factor by Tipping and Ogilvie \cite{tipping_1984}, which is also used in other models \cite{miller_phd_2012}.
\begin{align}
F_{rot}(J) = \left[1 + \kappa^2 \left(\frac{p_1}{p_0}\right) (J^2+3J+3)\right]^2
\label{eq:HW_factor}
\end{align}

\noindent where $\kappa=\sfrac{2B_e}{\omega_e}=1.695 \times 10^{-3}$ for N\textsubscript{2} with vibrational frequency $\omega_e$ = 2358.518 cm\textsuperscript{-1} \cite{palmer_carsft_1989}. The ratio between the first two coefficients of the anisotropic polarisability expansion is $\sfrac{p_1}{p_0}=3.168$ \cite{bohlin_2011}.

The polarisation anisotropy is dependent on the vibrational level according to \cite{asawaroengchai_1980}:
\begin{align}
\gamma'_v = \beta_e + \beta_e' r_e \left[\frac{3B_e}{\omega_e} + \frac{\alpha_e}{2B_e}\right] \left(v + \frac{1}{2}\right)
\label{eq:pol_anisotropy}
\end{align}

\noindent where for N\textsubscript{2}, $\beta_e=6.91 \times 10^{-25}$ cm\textsuperscript{3} is the equilibrium polarisation anisotropy, $\beta_e'=1.4 \times 10^{-16}$ cm\textsuperscript{2} is the first derivative of the polarisation anisotropy with respect to the internuclear distance, $r_e=1.098 \times 10^{-8}$ cm is the equilibrium internuclear distance and $\alpha_e=1.7305 \times 10^{-2}$ cm\textsuperscript{-1} \cite{palmer_carsft_1989} is the first rotational-vibrational interaction constant.

Finally, the population difference factor is calculated from a normalised Boltzmann distribution of rotational and vibrational states \cite{miller_phd_2012}:
\begin{align}
\Delta\rho_{v,J;v,J+2} &= N_{v,J} - \frac{2J+1}{2J+5}N_{v,J+2}\nonumber\\
N_{v,J} &= \frac{g_J (2J+1)}{Z} exp \left(\frac{-F_{v,J}}{k_B T}\right)
\label{eq:boltzmann}
\end{align}

\noindent where $Z$ is the partition function. For N\textsubscript{2}, if $J$ is even, the rotational-nuclear degeneracy is $g_J=6$ and if $J$ is odd, it is $g_J=3$. $F_{v,J}$ is the rotational energy of rotational state $J$ in vibrational level $v$. This energy is calculated based on the non-rigid rotor approximation (i.e. including centrifugal distortion) and taking into account the interaction between rotations and vibrations of the molecule \cite{herzberg_book_1989}:
\begin{align}
F_{v,J} &= \omega_e \left(v + \frac{1}{2}\right) - \omega_e\chi_e \left(v + \frac{1}{2}\right)^2 + BJ(J+1) - DJ(J+1)^2\nonumber\\
B &= B_e - \alpha_e \left(v + \frac{1}{2}\right) + \gamma_e \left(v + \frac{1}{2}\right)^2\nonumber\\
D &= D_e + \beta_e \left(v + \frac{1}{2}\right)
\label{eq:energy_levels}
\end{align}

\noindent where for N\textsubscript{2}, $D_e=5.774 \times 10^{-6}$ cm\textsuperscript{-1} and $\omega_e\chi_e=14.2935$ cm\textsuperscript{-1} are the equilibrium centrifugal constant and second order vibrational constant respectively. $\alpha_e=1.73035 \times 10^{-2}$ cm\textsuperscript{-1}, $\beta_e=1.55 \times 10^{-8}$ cm\textsuperscript{-1} and $\gamma_e=-3.1536099 \times 10^{-5}$ cm\textsuperscript{-1} are the rotational-vibrational interaction constants \cite{palmer_carsft_1989}. In our model the ground state ($v=0$) and first excited state ($v=1$) vibrational levels are included, as 3.5\% of the molecules are in the first excited vibrational state at 1000 K.

The transition frequencies for the S-branch, $\Delta E_{v,J;v,J+2}$, i.e. the energy differences between rotational levels $J+2$ and $J$ in vibrational level $v$, are calculated from Equation \ref{eq:energy_levels} according to the selection rules $\Delta v=0$ and $\Delta J=+2$ for S-branch transitions \cite{bohlin_phd_2012}:
\begin{align}
\Delta E_{v,J;v,J+2} &= F(v,J+2)-F(v,J)\nonumber\\
&= 2B(2J+3) - 4D(2J+3)(J^2+3J+3)
\label{eq:transition_frequency}
\end{align}

The transition linewidths, $\Gamma_{v,J;v,J+2}$, can be calculated from linewidths models, such as the modified exponential gap (MEG) model. Steinfeld \textit{et al.} \cite{steinfeld_1991} provide an overview of many of these linewidths models. However, the linewidths used in our computational code are based on interpolated experimental measurements by Kliewer \textit{et al.} \cite{kliewer_2012}.

The probe electric field in Equation \ref{eq:3rd_order_pol_specific}, $E_{pr} (t-\tau)$, is modelled as the sum of Gaussians, including a chirp factor \cite{yang_2017}:
\begin{align}
E_{pr}(t-\tau) &= E_{probe}(t-\tau) \times E_{chirp}(t-\tau)\nonumber\\
E_{probe}(t-\tau) &= \sum_{j} A_j exp \left[-\frac{2ln(2)(t-\tau_j)^2}{\tau^2_j}\right]\nonumber\\
E_{chirp}(t-\tau) &= exp \left[-i\frac{2ln(2)\alpha(t-\tau)^2}{\Delta t_{exp}^2}\right]
\label{eq:probe_beam}
\end{align}

\begin{figure}[htbp]
\centering
\includegraphics{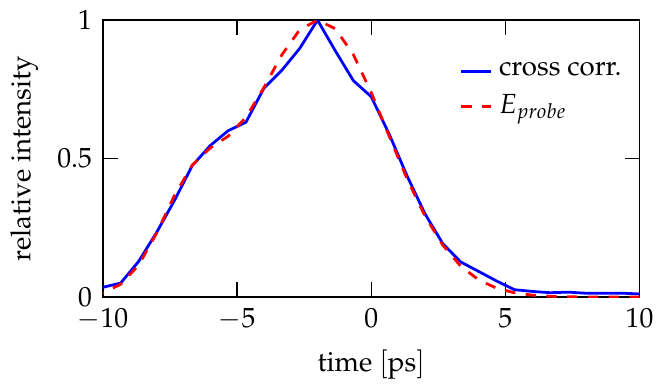}
\caption{Probe pulse in the time domain ($E_{probe}=\sqrt{I_{probe}}$) as measured by a cross-correlation (cross corr.) in argon (blue line). A two-Gaussian probe pulse (red line) as modelled by Equation \ref{eq:probe_beam} is shown as well.}
\label{fig:probe_pulse}
\end{figure}

A sum of two Gaussians ($j=2$) is chosen to fit measured cross-correlations of the probe beam in argon, which is not a perfect Gaussian, but shows significant asymmetry. Careful modelling of the probe pulse shape is required for accurate CARS spectra. The chirp factor, $\alpha$, is included in our model because we measured the bandwidth of the probe to be $\nu=3.0$ cm\textsuperscript{-1}. This bandwidth has a transform limit of 4.9 ps, but cross correlations in argon gave a pulse duration, $\Delta t_{exp}$, of 5.5 ps, 12\% longer than the TL. Including the chirp stretches the probe from its TL to the required length. Without this chirp factor, the model has narrower peaks and shifted frequencies when compared with experimental spectra. Further, including the chirp allows better modelling of spectrally overlapping transitions that can be seen in our experimental data. Figure \ref{fig:probe_pulse} shows a measured cross correlation of the probe beam, along with a two-Gaussian fit.

The final step in modelling CARS spectra is to convolve the spectrum with an instrument slit function, modelled as a 1.5 cm\textsuperscript{-1} Gaussian in our case. We determined this function by measuring the spectral linewidths of mercury emission lines near the probe frequency.

\begin{figure}[htbp]
\centering
\includegraphics{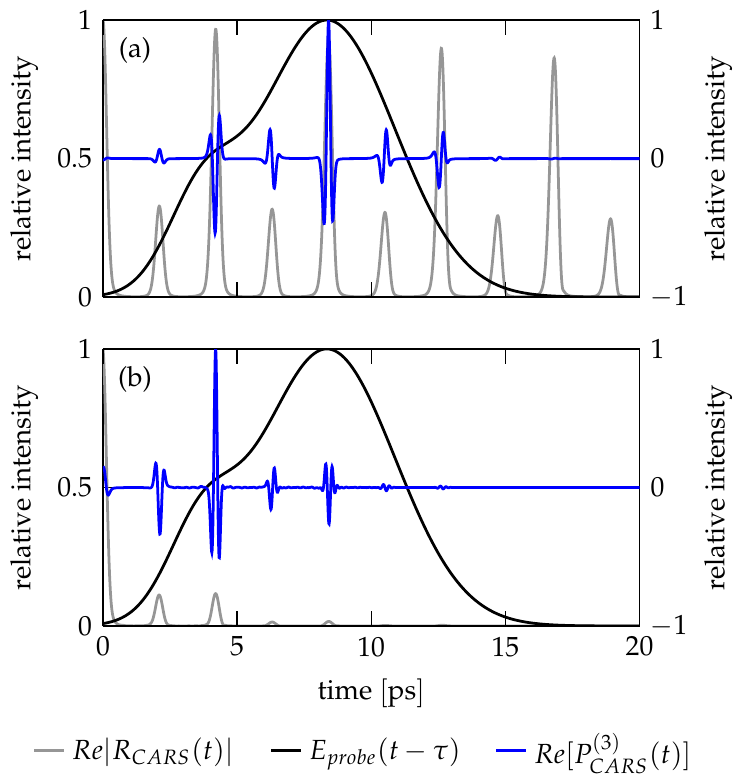}
\caption{Time domain model of the absolute value of the real part of the molecular response (same as Figure \ref{fig:N2_molecular_response}, but as an absolute value, grey line), the two-Gaussian probe pulse shape (black line) and the real part of the corresponding CARS signal (blue line). (a) displays these functions at 1 atm and (b) at 69 atm. In both cases, the model is for pure N\textsubscript{2} at 293 K and the peak of the probe pulse is centred at a time delay of 8.4 ps, the first full revival of the molecular response of N\textsubscript{2}. The probe pulse shape shown here is an example, based on one of the measured cross correlations.}
\label{fig:time_high_vs_low_P}
\end{figure}

Figure \ref{fig:time_high_vs_low_P} shows model results for two different conditions. Graph (a) shows the molecular response function, the probe pulse shape and the resulting CARS signal at 1 atm of N\textsubscript{2}, while (b) shows the same at a pressure of 69 atm. Looking at the molecular response (grey line, same as Figure \ref{fig:N2_molecular_response}, but as an absolute value), one can see that it decays very slowly at low pressure, but at high pressure it drops to less than a tenth within the first 10 ps, as the decay is exponential with pressure. However, there remains a significant CARS signal (blue line) even at high pressure, because the short picosecond probe (black line) overlaps with most of the molecular response. Additionally, the probe is not present at time zero, so it is not necessary to include the nonresonant signal in the model.

\begin{figure}[htbp]
\centering
\includegraphics{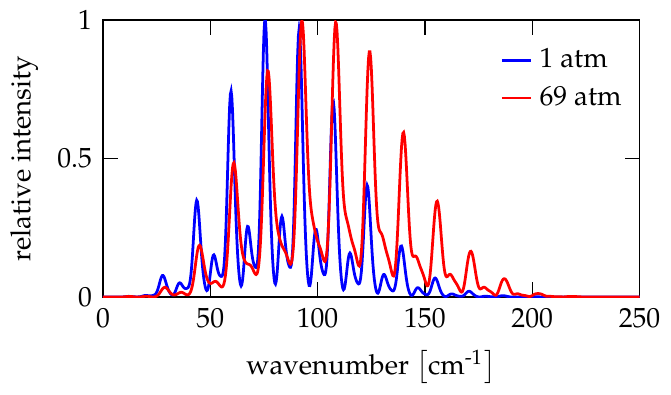}
\caption{Frequency domain N\textsubscript{2} HR-CARS spectra at 293 K and 1 atm (blue line) and 69 atm (red line), corresponding to the low and high pressure CARS signals in Figure \ref{fig:time_high_vs_low_P}. A 1.5 cm\textsuperscript{-1} slit function has been convolved with the spectra.}
\label{fig:freq_high_vs_low_P}
\end{figure}

In the frequency domain, the CARS spectra corresponding to the low and high pressure time-domain CARS signals of Figure \ref{fig:time_high_vs_low_P} are shown in Figure \ref{fig:freq_high_vs_low_P}. It can be seen that even at high pressure the lines are relatively narrow and well resolved, because the probe overlaps significantly with the molecular response. The non-centred shapes of the small lines (that arise from transitions involving odd rotational states) arise from the non-Gaussian shape of the probe pulse in combination with the chirp.

Our computational code can model the pump/Stokes beams explicitly, without assuming impulsive excitation. However, we found that the difference between the explicit and impulsive model is marginal, at least for the pump/Stokes beams used in our experiments. We therefore modelled our spectra with impulsive excitation as this is computationally faster. Modelling of the nonresonant response is unnecessary, as the spectra presented in this paper were delayed long enough so that no significant NR signal was present. The G-matrix, which accounts for line-mixing at very high pressures, was not included in the model. This is because according to our calculations, line mixing does not have a significant effect on rotational transitions, even at the highest pressures investigated (69 atm).

\section{Experimental Setup}
\begin{figure}[htbp]
\centering
	
\resizebox{1.0\columnwidth}{!}{%
\includegraphics{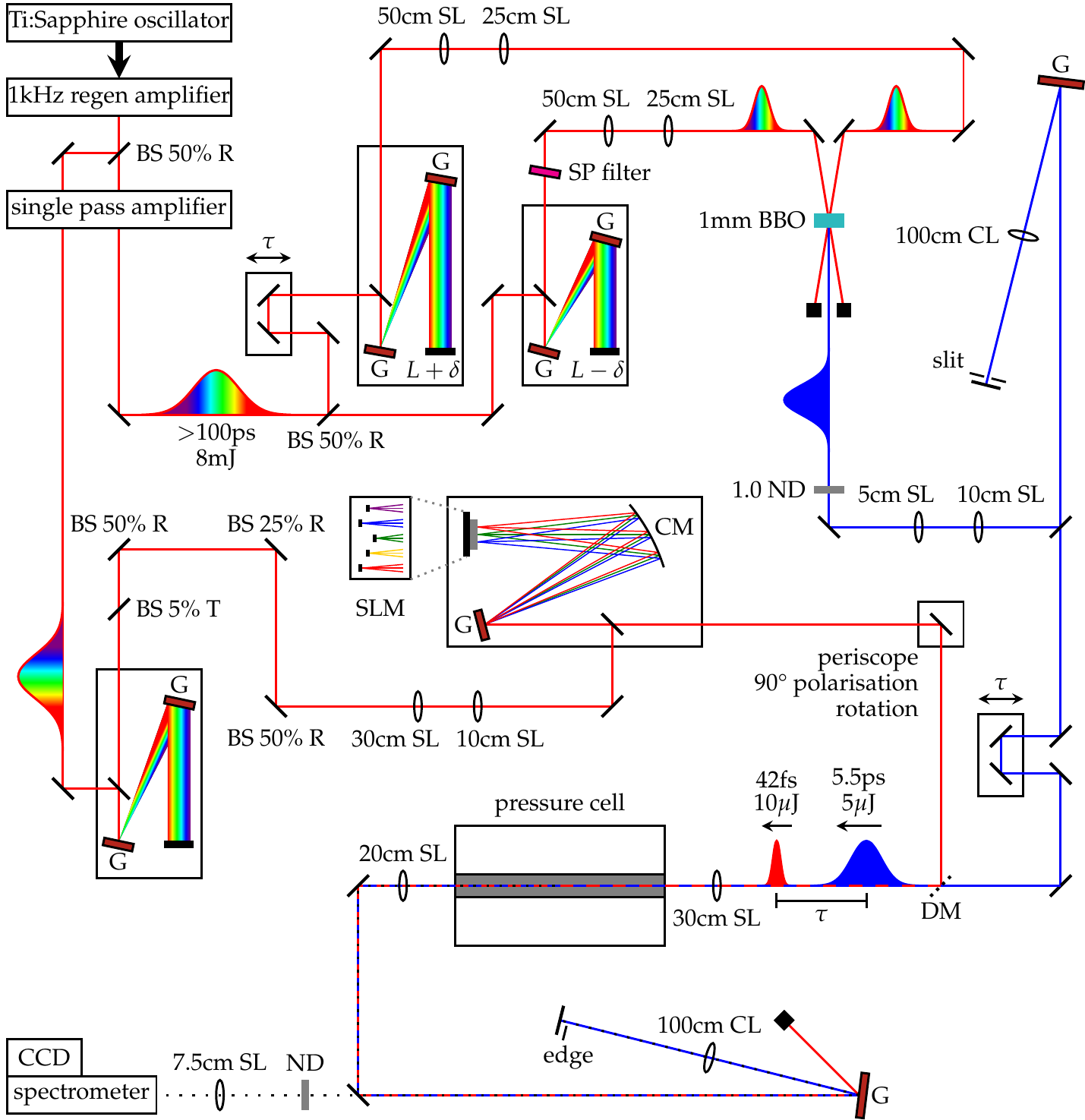}}
\caption{Two beam HR-CARS experimental setup. The beam from a 1 kHz Ti:Sapphire laser system is split into two equal power beams. The pump/Stokes pulse is generated by grating-compressing one of the output beams followed by dispersion compensation in a pulse shaper that uses a spatial light modulator (SLM). The probe is generated using second harmonic bandwidth compression (SHBC), by splitting the other amplifier output in two, chirping the two beams in grating compressors equally and opposite and combining them in a beta barium borate (BBO) crystal. Spectral wings are filtered in a folded 4\textit{f} grating filter. The pump/Stokes and probe beams are combined collinearly and focussed into a pressure cell. The resulting CARS signal is filtered in a folded 4\textit{f} grating filter and focussed into a spectrometer. R: reflection, T: transmission, SL: spherical lens, CL: cylindrical lens, G: grating, CM: curved mirror, DM: dichroic mirror, SP: short pulse, BS: beam splitter, ND: neutral density.}
\label{fig:complete_exp_setup}
\end{figure}

The experimental setup described here is the same as the one reported in a previous paper \cite{courtney_2019_a}. Both, the pump/Stokes and the probe beams, come from the same laser system, a Coherent Legend Elite Ti:Sapphire regenerative amplifier (790 nm, 1 kHz), that is seeded by a Coherent Vitara oscillator (100 MHz). The output is split into two beams of equal power before being single pass amplified to about 9 mJ/pulse. These resulting 795 nm broadband beams have a width of $\Delta\nu$ = 360 cm\textsuperscript{-1} and are strongly positively chirped ($>$100 ps). One of these output beams is used to create the femtosecond pump/Stokes beam and the other one is used to create the picosecond probe beam. A schematic diagram of the entire HR-CARS experimental setup is given in Figure \ref{fig:complete_exp_setup}.

\begin{figure}[htbp]
\centering

\resizebox{0.8\columnwidth}{!}{%
\includegraphics{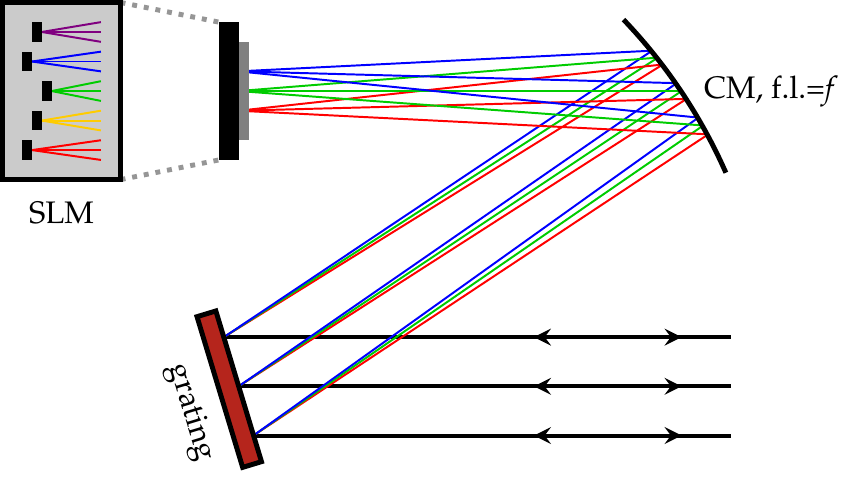}}
\caption{4\textit{f} pulse shaper. Light is dispersed in a grating and the different colours are focussed with a curved mirror onto a spatial light modulator (SLM), that is located at the Fourier plane. The liquid crystal SLM array can delay the different colours by different amounts of time, creating a nearly dispersion free pulse. CM=curved mirror.}
\label{fig:pulse_shaper}
\end{figure}

The pump/Stokes beam is created by using one of the stretched and amplified output beams and compressing it in a grating compressor to its near-transform limit of about 40 fs. Using a series of beam splitters, the energy of the beam is reduced to 35 $\mu$J/pulse. Furthermore, the beam is down-collimated using 300 mm and 100 mm spherical lenses, from a $\sfrac{1}{e^2}$ beam diameter of 12 mm to 4 mm. The beam then enters a pulse shaper (Femtojock). The reduction in power is dictated by the spatial light modulator (SLM) in the pulse shaper. A diagram of the pulse shaper is shown in Figure \ref{fig:pulse_shaper}. It uses a 4\textit{f} design to separate the different colours of the beam with a grating and focusses them with a curved mirror onto an SLM at the Fourier plane. The SLM is a liquid crystal array of pixels, to each of which a different voltage can be applied to create a relative delay of the different colours of the beam. The so-called multiphoton intrapulse interference phase scan (MIIPS) algorithm \cite{lozovoy_2004, xu_2006} is used to create a compensation phase mask to be applied to the SLM to account for dispersion in the pump/Stokes beam, mainly induced from the 28 mm high pressure cell fused silica entrance window. Linear and nonlinear dispersion are accounted for using this setup, creating a nearly transform-limited pulse at the interaction probe volume inside the pressure cell.

\begin{figure}[htbp]
\centering
\includegraphics{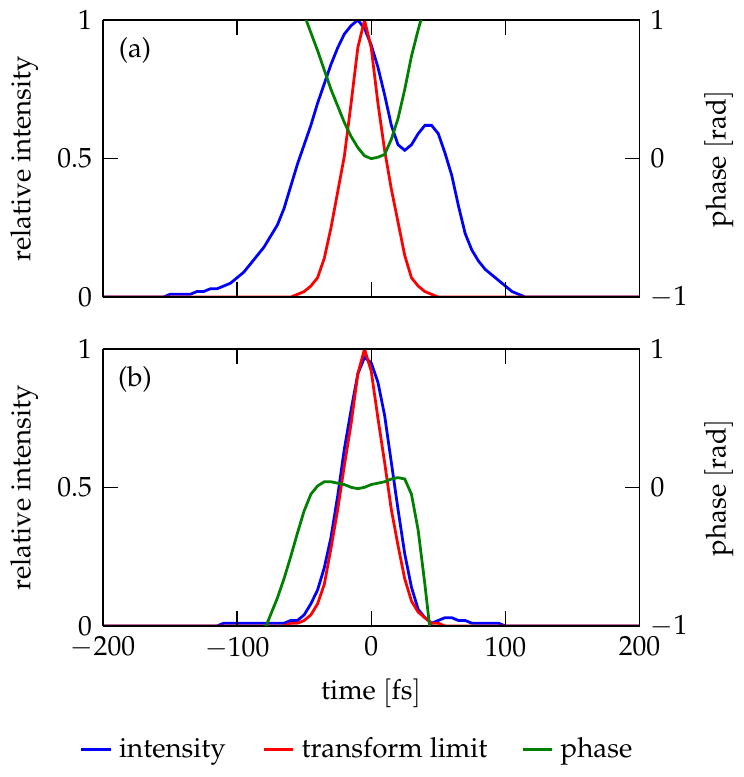}
\caption{FROG measured intensity (blue), transform limit (red) and phase (green) of the pump/Stokes pulse after passing first through the pulse shaper and then a 28 mm quartz window. (a) without a compensation phase mask: the intensity is far from its TL and has a width (FWHM) of 112 fs; the phase changes significantly over the length of the pulse. (b) with a compensation phase mask: the intensity is almost at its TL and has a width of 42 fs; the phase is almost constant over the length of the pulse.}
\label{fig:frog_mask_vs_nomask}
\end{figure}

Frequency-resolved optical gating (FROG) \cite{trebino_1993, trebino_1997} was used to measure the pulse to be 42 fs with an almost constant phase over the length of the pulse after passing through the pressure cell window, as shown in Figure \ref{fig:frog_mask_vs_nomask}. The energy of the beam after the pulse shaper is 12 $\mu$J/pulse, about a third of the input energy.

\begin{figure}[htbp]
\centering
\includegraphics{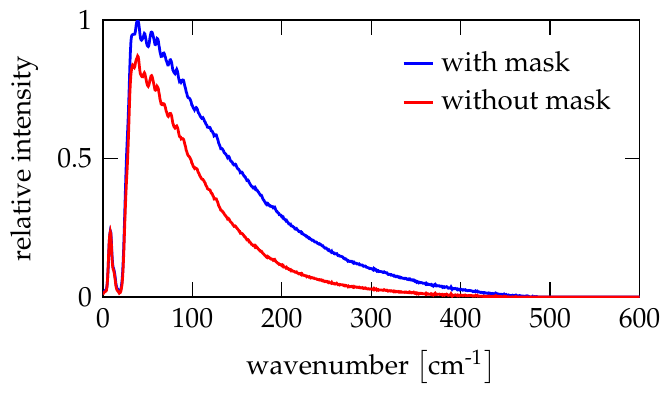}
\caption{Nonresonant CARS spectra of argon showing the pump/Stokes excitation efficiency. One is taken without the phase mask on the SLM (red line), corresponding to Figure \ref{fig:frog_mask_vs_nomask} (a) and one is taken with the phase mask on the SLM (blue line), corresponding to Figure \ref{fig:frog_mask_vs_nomask} (b).}
\label{fig:ar_mask_vs_nomask}
\end{figure}

Using the mask significantly increases the excitation efficiency of the pump/Stokes beam (also denoted the experimental spectral response). This can be seen in Figure \ref{fig:ar_mask_vs_nomask}, which shows two nonresonant argon CARS spectra, one recorded with the mask and one without. The excitation efficiency with the mask is better, as there is an increased NR response across all wavenumbers, compared to not using a mask. Especially at higher wavenumbers there remains a relatively large response when the mask is used. This is because pump/Stokes excitation at higher wavenumbers requires an increasingly larger frequency separation between intrapulse pump and Stokes interactions, which quickly lose time coincidence with chirp. NR argon spectra are used to correct experimental N\textsubscript{2} spectra for the pump/Stokes excitation efficiency.

The probe beam is generated from the other amplifier output beam via sum-frequency generation (SFG) using a second harmonic bandwidth compression (SHBC) method. This setup, adapted from a technique by Raoult \textit{et al.} \cite{raoult_1998}, is described in full detail in a different paper \cite{courtney_2019_b}. While most SHBC setups start with compressed beams \cite{yang_2017, kearney_2013}, our design feeds the uncompressed amplifier output to the SHBC setup, as can be seen in Figure \ref{fig:complete_exp_setup}. This reduces the number of gratings needed in the setup, making it simpler, while at the same time leading to a higher power throughput. The uncompressed strongly chirped amplifier output is split into two equally powerful beams, that are each routed through a grating compressor which have their gratings shifted in an equal but opposite way from the separation needed for complete pulse compression, thereby chirping the two beams equally but opposite. The two chirped beams are then down collimated with 500 mm and 250 mm spherical lenses from a beam size of 10 mm to 5 mm. This increases the energy density and allows for more energy-efficient sum-frequency generation, without damaging the crystal, as would happen if the beams were focused onto it. The two beams are combined in a 1 mm beta barium borate (BBO) crystal (Type1), in which their frequencies add via SFG to produce a narrowband unchirped picosecond pulse centred at 396.5 nm. However, in reality the chirp of the fundamental beams has nonlinear temporal contributions, resulting in a time-dependent SFG beam that is slightly blue-shifted from its minimum frequency at the edges of the pulse \cite{courtney_2019_b}.

\begin{figure}[htbp]
\centering
\includegraphics{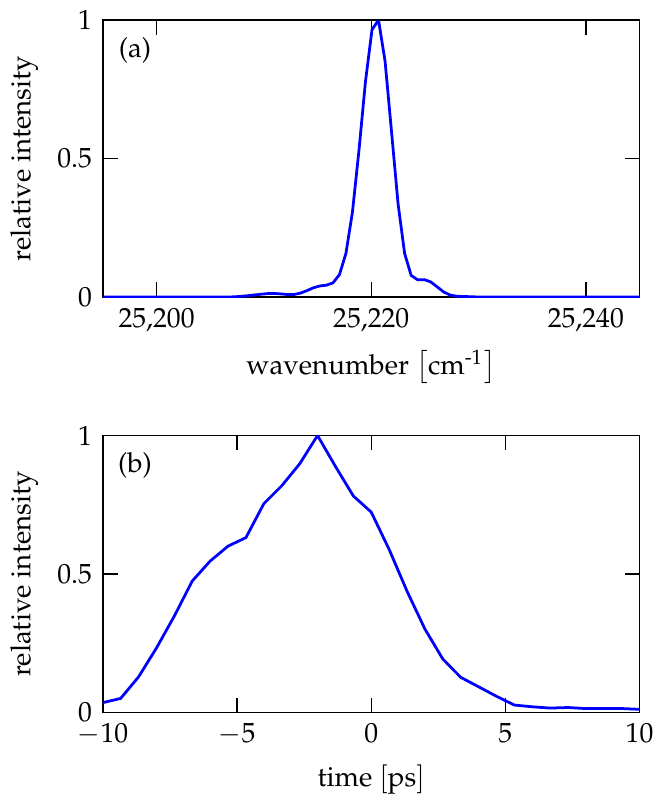}
\caption{Probe beam as created by the SHBC setup described above. (a) in the frequency domain measured in a spectrometer, FWHM of 3.5 cm\textsuperscript{-1} (3 cm\textsuperscript{-1} after deconvolution with the spectrometer's 1.5 cm\textsuperscript{-1} slit function). (b) in the time domain as measured with a cross correlation in nonresonant argon.}
\label{fig:probe_beam}
\end{figure}

The beams are crossed at a very narrow angle ($\sim 1^\circ$), to minimise spatial chirp. The resulting beam has a full width at half maximum (FWHM) bandwidth of 3 cm\textsuperscript{-1} (deconvolved with the spectrometer’s slit function, separately measured to be 1.5 cm\textsuperscript{-1}) and a duration of about 5.5 ps, as confirmed with a cross-correlation of the NR signal in pure argon. Figure \ref{fig:probe_beam} shows these beam characteristics in the frequency and time domain. A translation stage in one of the fundamental beams before the BBO crystal is used to control the relative delay between the oppositely chirped pulses, allowing for fine tuning of the SFG spectrum. Further, two stacked angle-tuned short pass filters in the other fundamental beam are used to cut out the leading edge of that pulse, resulting in an SFG pulse with a relatively clean and steep early time profile, albeit not as steep as anticipated. This SHBC setup has an energy conversion efficiency of 20\% as the combined 6 mJ/pulse from the input pulses result in a 1.2 mJ/pulse SFG beam. The SFG beam emerging from the BBO crystal is filtered with a neutral density filter (1.0 ND) to reduce its power to avoid damaging the grating of the grating filter (see below) and collimated with 50 mm and 100 mm spherical lenses, to reduce spatial variations in the beam intensity. At this point, the beam has spectral wings arising from imperfect conjugate chirping (frequency distributions at each point in time in the broadband pulses), together with blue-shifted satellite peaks that are a result of interferences in the SFG of temporally chirped fundamental pulses \cite{courtney_2019_b}. These are filtered out in a folded 4\textit{f} grating filter (f.l.=1 m, 1800 grooves/mm grating) with a slit at the Fourier plane. The CARS signal lineshapes are significantly simplified when the satellite pulses are filtered out. The spectral wings, though a lot weaker than the peak of the probe, are similar in strength to the CARS signal. Their reduction diminishes the background noise coming from the probe, that otherwise is present in collinear CARS.

The two beams are routed to a high pressure gas cell in which they are combined collinearly through the use of a dichroic mirror that transmits the 396.5 nm probe beam and reflects the 800 nm pump/Stokes beam. Before combination in the dichroic mirror, the polarisation of the pump/Stokes beam is rotated by 90$^\circ$ in a periscope, so that the two beams have matched vertical polarisations. Time delay between the pulses is provided by a computer-controlled translation stage (ThorLabs NRT100). The beams are focussed into the cell with an uncoated spherical lens of focal length 300 mm. The power of the beams just before the cell is measured to be 10 $\mu$J/pulse for the pump/Stokes pulse and 5 $\mu$J/pulse for the probe beam. The $\sfrac{1}{e^2}$ diameters at the beam waists are 100 $\mu$m and 70 $\mu$m respectively. The collinear geometry is chosen to achieve an increased signal from the rather low power input beams by creating a longer CARS interaction region, which is still much shorter than the length of the cell. The custom-built pressure cell itself is an insulated steel cylinder of length 30 cm and diameter 15 cm, with the gas to be probed being contained in a core of diameter 2.5 cm in the centre of the cylinder. Self-phase modulation of the pump/Stokes beam that has been reported for other CARS experiments \cite{gu_2019} was not observed, due to the low power of the beam. After the cell, a 200 mm focal length spherical lens is used to collimate the exiting pump/Stokes, probe and signal beams. To filter the probe from the signal beam, a 4\textit{f} grating filter with a sharp edge at the Fourier plane is used. This ensures that the probe is almost completely removed from the signal, while retaining all CARS transitions, including even the $J=2$ lines, which are just 28 cm\textsuperscript{-1} from the probe. Optical short-pass filters are not able to provide a sufficiently sharp edge, which is why the grating filter is used. A 75 mm spherical lens is used to focus the CARS signal onto the entrance of a 0.550 m Czerny-Turner spectrometer (Horiba iHR550, 2400 grooves/mm grating). It is then dispersed onto a back-illuminated CCD (Andor DU971N-BV, 400 $\times$ 1600 pixel, 16 $\mu$m pixel size). 1 kHz single-shot data collection is achieved by vertically binning the CARS signal in an 11 $\times$ 1000 pixel region in $<$1 ms (10 $\mu$s exposure).

\begin{figure}[htbp]
\centering
\includegraphics{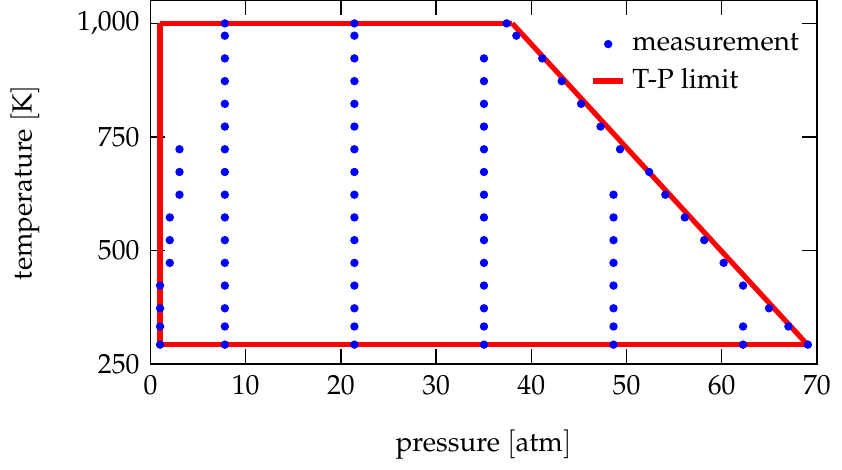}
\caption{Maximum and minimum temperatures and pressures for which the pressure cell is rated (bound by the red trapezoid) and temperatures and pressures at which HR-CARS spectra were recorded (blue dots).}
\label{fig:tp_trapezoid}
\end{figure}

N\textsubscript{2} HR-CARS spectra were measured over a large range of temperatures and pressures. The high pressure stainless steel cell is rated for up to 70 atm at room temperature, with the maximum pressure decreasing linearly with increasing temperature, to 38 atm at 1000 K. This leads to a trapezoid of temperatures and pressures (T-P) that can be achieved with this cell, as shown in Figure \ref{fig:tp_trapezoid}, which also shows all the temperatures and pressures at which spectra were taken. At each of those T-P points, several spectra were taken at different time delays, all 1 ps apart and centred at about 8.4 ps, which is the time of the first full revival of N\textsubscript{2}. Temperatures were measured with a thermocouple inserted into a separate hole in the steel cylinder, about 5 cm from the centre of the cell. The thermocouple measurements were validated by separately testing the cell with an additional thermocouple that was placed in the core of the cell, which holds the probed gas. It was found that the readings for both thermocouples were in good agreement. To ensure that the gas had the same temperature as the cell, the cell was kept sealed for some minutes prior to measurement or kept at a minimum flow necessary for maintaining the desired pressure. Nonresonant argon spectra were also recorded, as part of cross correlations to determine the probe pulse shape in time, and also to correct for the excitation efficiency of the pump/Stokes beam. At each temperature and pressure, 1000 single laser shots were collected.

\section{Fitting Procedure}
\begin{figure}[htbp]
\centering
\includegraphics{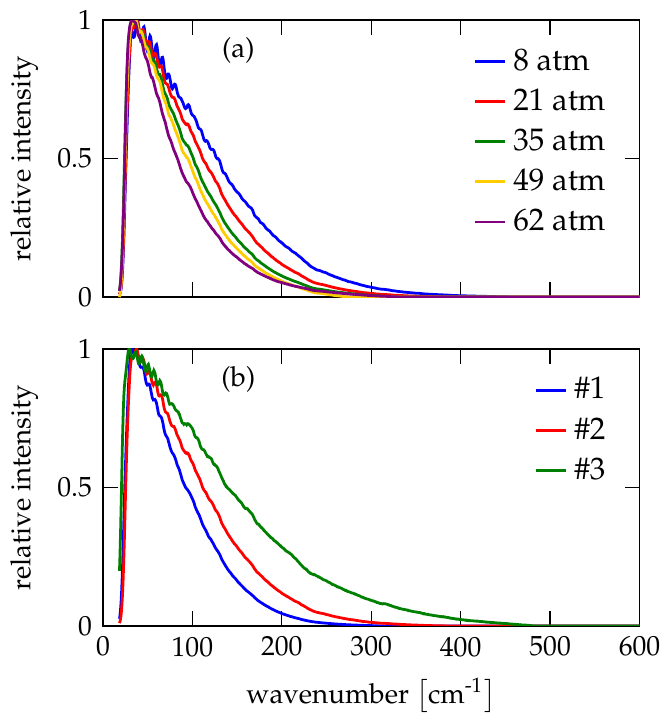}
\caption{Nonresonant CARS spectra of argon. (a) displays spectra taken at the same time but different pressures. (b) displays spectra taken at the same pressure (21 atm) but at different times during the same day. The time elapsed between spectrum \#1 and \#2 is 2:18 h and between \#2 and \#3 is 1:41 h.}
\label{fig:AR_change}
\end{figure}

In preparation for fitting the experimental spectra to the computational model, the 1000 single-shot spectra that were recorded at each temperature, pressure and time delay were averaged and background subtracted. Further, they were divided by a background-subtracted nonresonant argon spectrum to correct for the pump/Stokes excitation efficiency. Finally, the experimental CARS spectra were normalised. The argon spectra used for correction were taken on the same day and at the same pressure as the CARS spectra. This is because we observed narrowing of NR argon spectra with pressure and in addition, the NR spectra changed with time, from day to day and even during the same day. This is shown in Figure \ref{fig:AR_change}, which displays a clear trend in narrowing of NR spectra with increasing pressure and also shows significant differences between NR spectra taken at the same pressure but different times. The narrowing with pressure can be explained by the increasing number density of molecules inside the cell that the beam has to travel through before the probe volume and which induces chirp. The variation of the nonresonant response with time is believed to arise from the pulse shaper; slight alignment drift over time (common to many laser systems) into the SLM can have large effects on the dispersion of the pump/Stokes pulse after passing through the pressure cell window, which in turn has a big effect on the excitation efficiency. This shows how important it is to measure NR argon spectra close in time to the real CARS spectra and at the same pressure, as otherwise the temperature fits will be off by a significant amount if a wrong pump/Stokes excitation efficiency correction is used.

To obtain a wavenumber axis for the experimental spectra, a linear fit of position of the experimental CARS peaks in pixel space to their theoretical position in wavelength space is made using a low pressure CARS spectrum with a time delay close to the first revival of N\textsubscript{2}, because of its simplicity and high resolution. For this, the theoretical transition frequencies are calculated in wavenumbers. The wavenumber of the probe beam (25221 cm\textsuperscript{-1}) is added and the result is converted to wavelength. These numbers are then fit linearly to their positions in pixel space. This leads to a linear conversion equation of the form $wavelength = m \times pixel + c$, where $m$ is the gradient and $c$ the intercept. To represent experimental spectra on a wavenumber axis, the calibrated wavelength axis is inverted to wavenumber and the wavenumber of the probe is subtracted. The gradient and intercept values for different spectra at different temperatures were all very close, on average $m=-0.009569$ nm/pixel (standard deviation of 7.2$\times$10\textsuperscript{-6}) and $c=396.238$ nm (standard deviation of 0.016) respectively. Regression statistics give an average R\textsuperscript{2} value of 0.99999 showing that a linear wavelength calibration is satisfactory.

Experimental spectra are then fit to the computational spectra through a non-linear least squares fitting routine in MATLAB called \textit{lsqcurvefit}. This allows one to define fitting parameters and to place lower and upper constraints on those parameters. In our fitting, the wavenumber axis was allowed to slightly shift and stretch by allowing the gradient ($m$) and intercept ($c$) of the linear wavelength calibration to vary within three standard deviations. The relative height of the model to the experimental spectra was varied between 95\% and 105\% to account for a possible intensity offset of the highest peak. Because of the non-Gaussian shape of the probe pulse and the resulting uncertainty in exact probe pulse delay, the temporal centre of the probe pulse was allowed to vary within $\pm$2 ps of the experimentally verified delay based on the corresponding cross correlation. This was done for the most resolved spectrum at each temperature and pressure. The time delays for the other spectra at a given temperature and pressure are then steps of 1 ps $\pm$0.2 ps from the best fit probe delay of the most resolved spectrum. Similarly, the chirp parameter, $\alpha$ in Equation \ref{eq:probe_beam}, was floated between 0 and -2.0 for the most resolved spectrum and its best fit value $\pm$0.2 was used for the other time delays at the same temperature and pressure. Finally, the temperature was allowed to vary within $\pm$30\% from the thermocouple temperature reading recorded for a given spectrum. The pressure was kept fixed at its measured value.

\begin{figure}[htbp]
\centering
\includegraphics{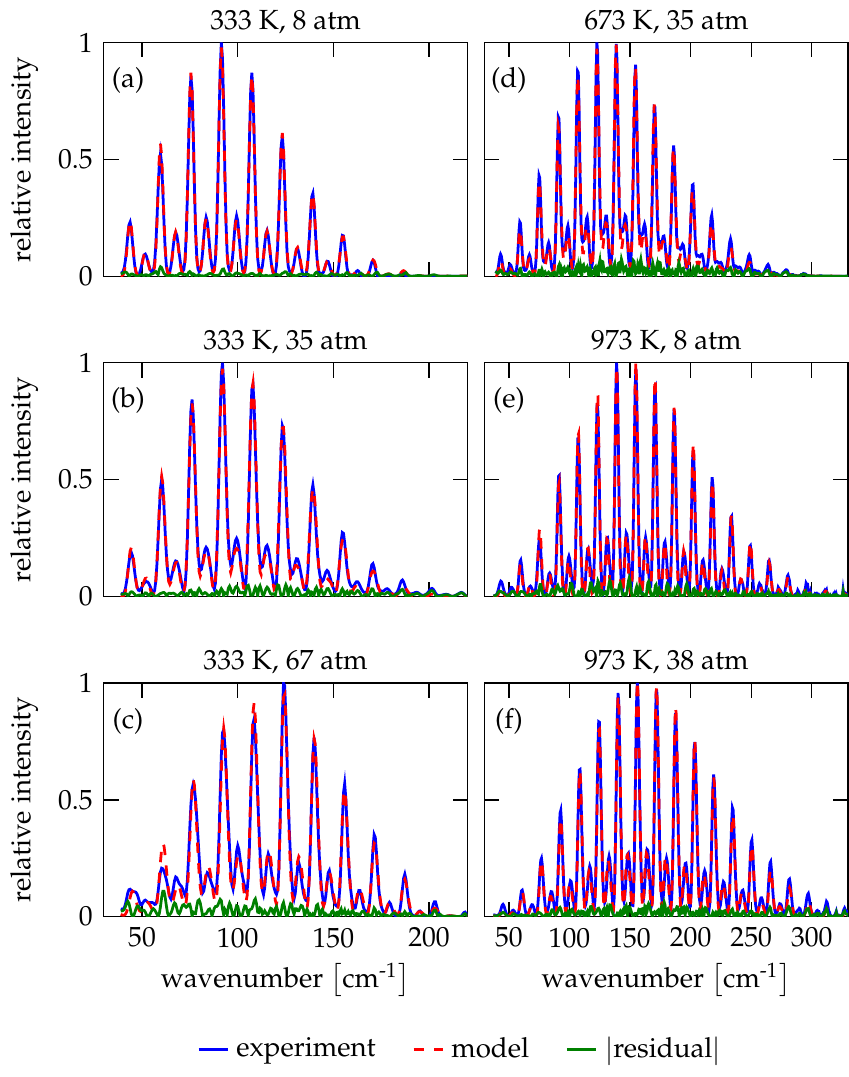}
\caption{Experimental and modelled HR-CARS spectra of N\textsubscript{2}. Experimental (blue line) and best-fit model spectra (dashed red line) are shown together with the absolute value of the residual (experiment-model, green line). The spectra were all taken at a probe delay of 8 ps, apart from (c), which was taken at a delay of 11 ps. The best fit temperatures and percentage differences to the corresponding thermocouple measurements (as a percentage of the thermocouple measurement) are: (a) 341 K, 2.3\%; (b) 322 K, -3.2\%; (c) 312 K, -6.2\%; (d) 685 K, 0.4\%; (e) 980 K, 1.1\%; (f) 955 K, -1.8\%.}
\label{fig:expt_vs_model}
\end{figure}

All spectra were fit according to this procedure. Six fits of spectra taken at very different temperatures and pressures are shown as an example in Figure \ref{fig:expt_vs_model}. The spectra shown were all taken at a probe delay of 8 ps, apart from the highest pressure one, (c), which was taken at a delay of 11 ps, as at this pressure the earlier probe time delay spectra have a slight nonresonant contribution due to the resonant CARS signal being relatively low at very high pressure. It can be seen that across all the different temperatures and pressures fits are generally very good, as confirmed by the low residuals. Further, the best fit temperatures of the shown spectra are accurate, the percentage differences between best fit temperatures and thermocouple readings (as a percentage of the thermocouple reading) range from -6.2\% to 2.3\%.

\begin{figure}[htbp]
\centering
\includegraphics{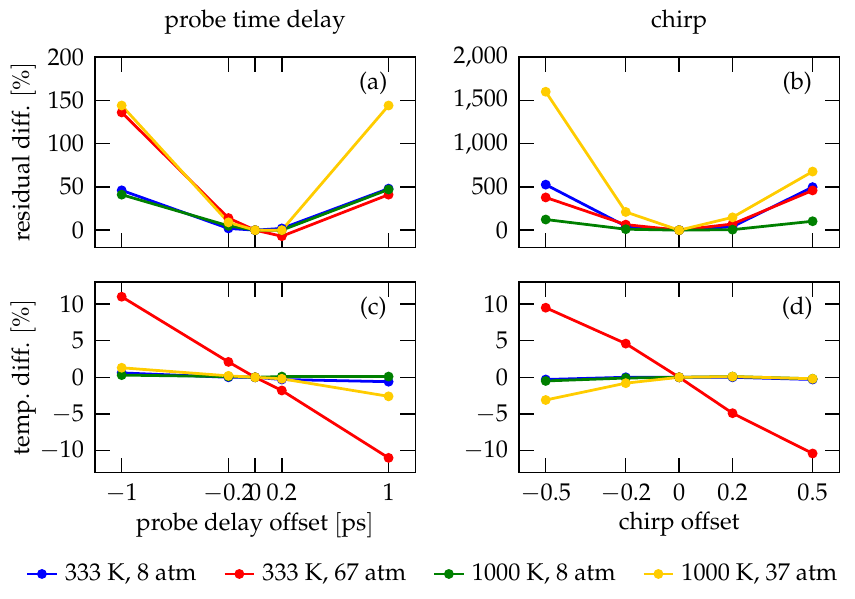}
\caption{Sensitivity study of probe delay and chirp. (a) and (b) show the percentage differences in residual as compared to the best fit residual (residual diff.) for different probe delay and chirp offsets from their best fit. (c) and (d) show the corresponding percentage differences in temperature as compared to the best fit temperature (temp diff).}
\label{fig:sensitivity_study}
\end{figure}

A study of the sensitivity of the fits on the chirp and probe delay was performed on four spectra from the four corners of the T-P trapezoid (Figure \ref{fig:tp_trapezoid}) at 333 K and 8 atm, 333K and 67 atm, 1000 K and 8 atm, 1000 K and 37 atm. This sensitivity study is not statistically representative of all spectra taken, but should serve as a strong indication as to how much the chirp and time delay affect the best fit. For this study, the chirp and probe delay were changed in turn, while all the other fitting parameters were kept at their best fit value, as found by the afore described fitting procedure. The chirp parameter, $\alpha$, was offset by $\pm$0.2 and $\pm$0.5 from its best fit value and the probe time delay was offset by $\pm$0.2 ps and $\pm$1.0 ps. In each of these cases, the residual and temperature fit were recorded. Figure \ref{fig:sensitivity_study} shows the results as a percentage difference of the residual and temperature, relative to their best fit values.

For both, chirp and probe delay, it can be seen that there is a significant increase in residual when these two parameters are changed from their best fit value, especially when the change is large but still within their uncertainty. However, the temperature fits do not follow this trend. For all tested spectra apart from the highest pressure one (67 atm) the temperature fits only change by a few percent, even if the corresponding residual increases by more than 100\%. Only the highest pressure spectrum shows changes of more than 10\% in the temperature fits. Overall, varying the fitting parameters significantly increases the residual, but does not affect the temperature fits significantly, except for the highest pressure spectra. This shows that the high pressure spectra need to be modelled very carefully to get a good fit with the experiment.

\section{Best Fit Temperatures}
All spectra taken at the various pressures, temperatures and time delays were fit using the afore described fitting procedure. Across all spectra, the average percentage temperature difference of the best fit temperature to the thermocouple temperature (as a percentage of the thermocouple reading) is -0.3\% with a standard deviation of 7.1\%. This means that 84\% of the fits fall within $\pm$10\% of the measured thermocouple temperature. The average absolute percentage temperature difference (opposed to the average percentage temperature difference, which averages positive and negative percentage temperature differences) is 5.3\%.

\begin{figure}[htbp]
\centering
\includegraphics{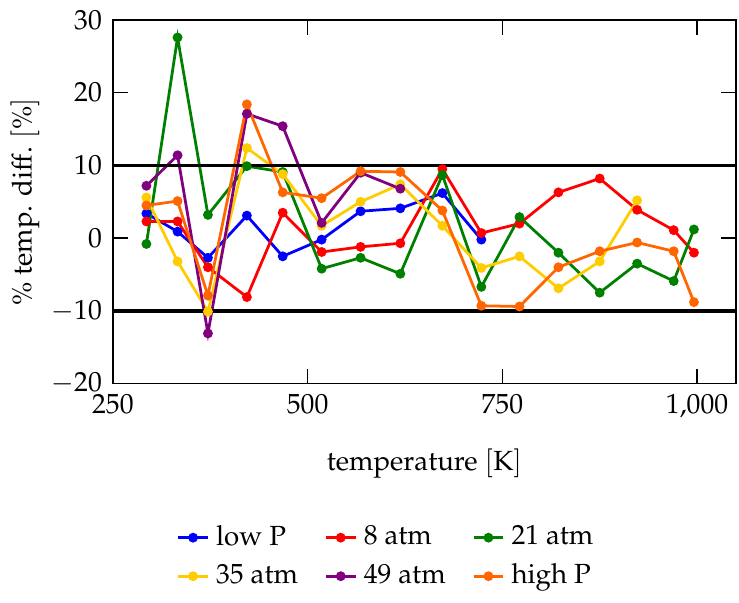}
\caption{Plots of \% temperature difference (temp. diff.) between the measured thermocouple temperature and the best fit temperature vs. absolute measured temperature. Each line shows data at a different pressure. The black lines, shown for reference, indicate the $\pm$10\% range. See text for definition of \emph{low P} and \emph{high P}.}
\label{fig:results_8ps}
\end{figure}

We focus a statistical analysis on the data taken at a time delay of 8 ps, the time delay which is closest to the first full revival of N\textsubscript{2} and generally results in the most well resolved spectra. Best fit temperature differences to the thermocouple reading at all the tested pressures and temperatures for this time delay are shown in Figure \ref{fig:results_8ps}. Each line in this graph represents a different pressure. Note that the high pressure line contains the data for the highest pressure measured at each temperature, starting at 69 atm at 293 K and decreasing linearly to 37 atm at 1000 K, following the inclined edge of the T-P trapezoid (Figure \ref{fig:tp_trapezoid}). Similarly, the low pressure data follow the left edge of the T-P trapezoid. At low pressure and 8 atm the results are most accurate, with more than 95\% of the data lying between $\pm$10\% of the measured thermocouple temperatures. For medium pressures (21 atm and 35 atm) that number drops to 73\% and 87\% respectively. But even at the highest pressures, 78\% of the data lie within $\pm$10\% of the thermocouple measurement, indicating a good accuracy.

It should be noted that the low pressure best fit temperatures (Figure \ref{fig:results_8ps}, blue line) mostly lie within $\pm$5\% from the thermocouple temperature, not quite as accurate as previously reported accuracies, which are about 1\% for rotational N\textsubscript{2} spectra at atmospheric pressure \cite{miller_2011_b}. This is because the experimental setup is optimised for high pressure CARS experiments, and some of its components (e.g. the high pressure cell windows and the pulse shaper) add complexity to the setup making the experimental conditions less ideal than they could be for purely low pressure CARS studies. However, this complexity is considered necessary for the acquisition of well-resolved and relatively strong high pressure CARS spectra, thereby helping to improve the accuracy at higher pressures.

\begin{figure}[htbp]
\centering
\includegraphics{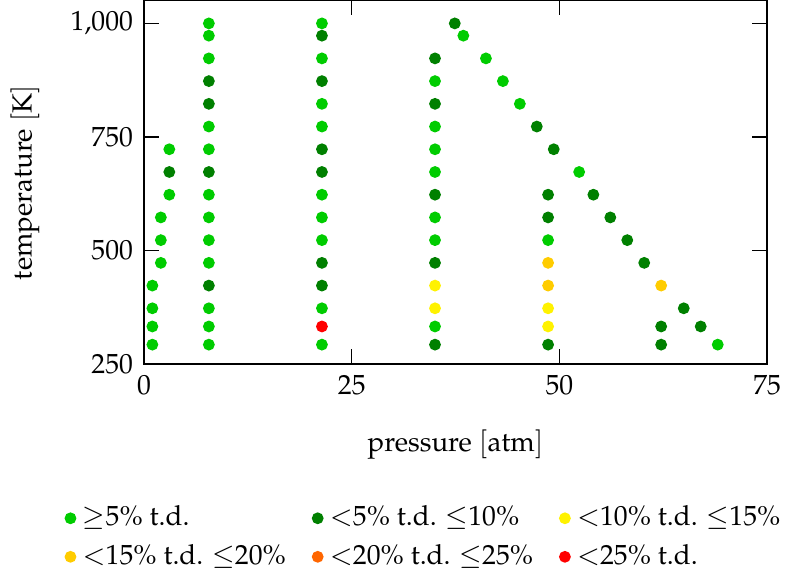}
\caption{Percentage temperature difference (t.d.) for 8 ps probe delay spectra on a temperature-pressure graph. The fits are grouped such that all data within the same 5\% accuracy interval (percentage difference between best fit and thermocouple-measured temperature) have the same colour.}
\label{fig:results_TP_8ps}
\end{figure}

To get a better overall understanding of the best fit data, a plot of all the fits at 8 ps probe delay is shown on a T-P trapezoid, in Figure \ref{fig:results_TP_8ps}. The fits are grouped such that all data within the same 5\% accuracy interval have the same colour. It can be seen that the majority of the fits are of high accuracy, across all temperatures and pressures. Especially at low pressures, the fits are all very accurate, the same is also true for fits at higher temperatures across all pressures. It is only at the higher pressures at low temperatures that some fits are less accurate than desired.

\begin{figure}[htbp]
\centering
\includegraphics{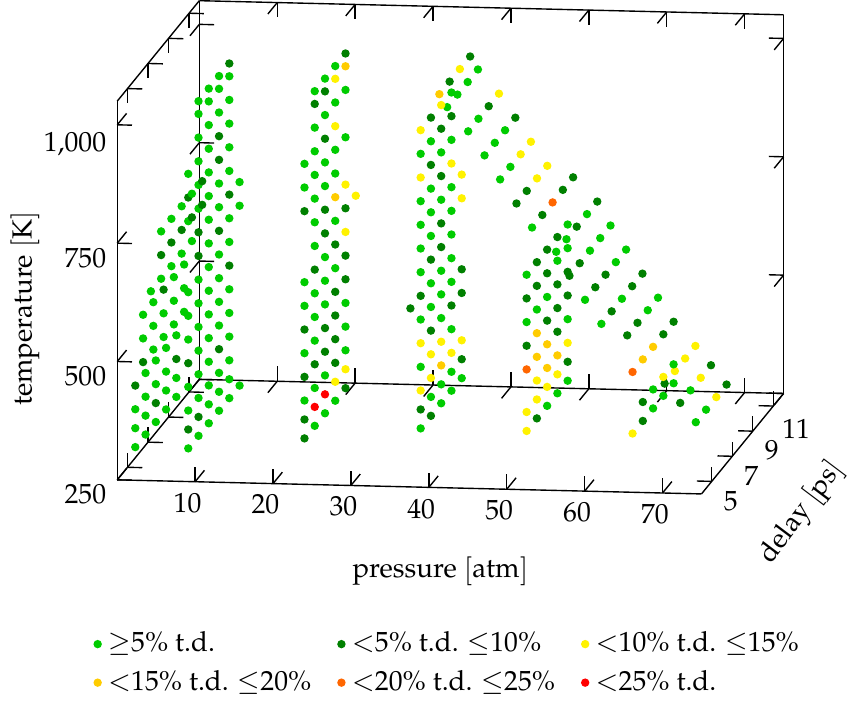}
\caption{Percentage temperature difference (t.d.) of all recorded spectra on a temperature-pressure graph. The fits are grouped such that all data within the same 5\% accuracy interval (percentage difference between best fit and thermocouple-measured temperature) have the same colour.}
\label{fig:results_TP_full}
\end{figure}

When looking at the T-P trapezoid of all the different time delays, the results look similar (Figure \ref{fig:results_TP_full}). At low pressures (up to 8 atm) the fits are all very accurate, across all temperatures. Also at medium pressures (between 20 and 40 atm), most fits are within $\pm$10\% of the measured thermocouple temperature, especially at higher temperatures. Even at the highest pressures (above 50 atm), most fits are still of acceptable accuracy. However, in this pressure regime, there are also a few fits that are less accurate. On average, however, the accuracy is good both a low and at high pressures, while the precision is a lot better at lower temperatures. At 8 atm, the average percentage difference between the best fit temperature and the measured thermocouple temperature, across all temperatures and time delays is 0.4\% (absolute average: 3.0\%) with a standard deviation of just 3.9\%. For the high pressure data, the average is a relatively small -2.2\% (absolute average: 7.0\%) with a larger standard deviation of 8.4\%.

Some of the larger deviations of the best fit to the measured thermocouple temperatures can be explained by a drift of the pump/Stokes beam pointing and dispersion with time. This especially affects the effectiveness of the pulse shaper to create an almost TL beam. The usefulness of the pulse shaper to create an almost TL pulse after passing through the window of the pressure cell (Figure \ref{fig:frog_mask_vs_nomask}), and therefore to generate a larger pump/Stokes excitation bandwidth (Figure \ref{fig:ar_mask_vs_nomask}) has been discussed above. But this relies on perfect alignment onto the SLM of the pulse shaper. Any slight alignment deviations that occur in time, which are common for many laser systems, will result in a considerable change in the pump/Stokes excitation bandwidth (Figure \ref{fig:AR_change}). If the excitation bandwidth changes between the measurement of a CARS spectrum and its corresponding NR Ar spectrum, then the division by the NR spectrum to correct for the pump/Stokes excitation bandwidth will result in an incorrect best fit temperature, leading to best fit temperature outliers. This highlights the need for taking many NR Ar spectra throughout an experimental run when using a pulse shaper, so that the pump/Stokes excitation bandwidth is the same for a CARS spectrum and its corresponding NR argon spectrum.

Pump/Stokes beam alignment drift should affect low and high pressure spectra similarly. But from the full set of results (Figure \ref{fig:results_TP_full}), it can be seen that most of the less accurate best fit temperatures occur at higher pressures. One explanation for this is that while the pump/Stokes excitation differences due to alignment drift affect all spectra equally, the NR spectra start to narrow with increasing pressure, as discussed before (Figure \ref{fig:AR_change}). There could be a variation in the degree of narrowing depending on the initial pump/Stokes excitation bandwidth at atmospheric pressure. This could lead to a compounding of the two effects (different pump/Stokes excitation bandwidth between the N\textsubscript{2} CARS and NR Ar spectrum and the narrowing of the excitation bandwidth with increasing pressure), leading to more inaccurate high pressure spectra than low pressure spectra, while maintaining similar average accuracies, as the effect is random. This is further amplified by high pressure spectra, in general, having a lower signal-to-noise ratio, resulting in less accurate fits. This decrease in signal strength at high pressures is due to the time domain molecular response only increasing linearly with number density, but decreasing exponentially with pressure, leading to an overall decrease in the signal (see Figure \ref{fig:time_high_vs_low_P}). Unfortunately, it is not possible to use a pulse shaper compensation mask at high pressure, as there is no way to get high pressure feedback signal to the pulse shaper algorithm. A potential solution for a chirp precompensation of the pump/Stokes pulse that is more stable with time could be the use of chirped mirrors. These mirrors induce a larger group delay dispersion to longer wavelengths than to shorter wavelengths. This precompensates a beam for the chirp induced while passing through an optical system, for example fused silica windows. However, chirped mirrors only correct for linear chirp, while the pulse shaper corrects for non-linear chirp as well.

Another factor that could contribute to the lower accuracy at higher pressures is uncertainty in the experimentally measured linewidths that are used in the model. Small errors in the linewidths have an increasingly larger effect on the model spectrum (and hence the best fit temperature), with increasing pressure, due to the exponential dependence of the CARS signal on the linewidths (see Equation \ref{eq:molecular_response}).

\section{Conclusion}
In conclusion, we have developed a hybrid rotational fs/ps CARS experimental setup that allows the collection of clean and well resolved spectra over a wide range of temperatures and pressures, in previously unexplored T-P regimes. This has been done using a pulse shaper to create a near TL pump/Stokes pulse at the interaction volume and by making a 5.5 ps probe beam in an SHBC setup that allows probing of the molecular response close to but not overlapping the nonresonant response. Further, we have developed a computational model that generates theoretical CARS spectra and can be used for fitting the experimental spectra to infer temperature from the spectral structure. While being based on a few assumptions and simplifications, much of the complexity of the CARS theory is retained in the model. The temperature fits reveal that our model is relatively accurate with 84\% of all acquired experimental spectra fitting with an error of 10\% or less. Even though more accurate at lower pressures, the model still gives good fits at high pressures, both qualitatively and quantitatively. However, it was also found that there are some less accurate fits. This should be taken into consideration for future experiments.

Now that we have shown that accurate thermometry is possible with HR-CARS at the high temperatures and pressures encountered in real combustion environments, the next step will be to further develop and test this technique to achieve even higher accuracy of temperature measurements. One day, this experiment could be performed in a real optical engine, to shed light on the robustness and suitability of this technique for temperature measurements in turbulent systems.\\

This material is based upon the work supported by an award from the United States Department of Energy’s Office of Science Early Career Research Program, which supported the development of the CARS model, the laser equipment, and C.J.K. The construction of the high pressure cell flow reactor, the femtosecond pulse shaper, and T.L.C. was supported by the Laboratory Directed Research and Development (LDRD) program at Sandia National Laboratories, which is a multimission laboratory managed and operated by National Technology and Engineering Solutions of Sandia, LLC., a wholly owned subsidiary of Honeywell International, Inc., for the U.S. Department of Energy’s National Nuclear Security Administration under Contract No. DE-NA0003525. The views expressed in this article do not necessarily represent the views of the U.S. Department of Energy or the United States Government.\\

B. Peterson gratefully acknowledges financial support from the European Research Council (ERC grant \#759546).\\

\noindent\textbf{Disclosures.} The authors declare no conflicts of interest.

\bibliography{references}

\end{document}